\documentclass[letter]{aa}  

\pdfoutput=1
\usepackage{graphicx}
\usepackage{array,multirow,makecell}
\usepackage{txfonts}
\usepackage{natbib}
\usepackage{amsmath}
\usepackage{ulem} 
\bibpunct{(}{)}{;}{a}{}{,} % to follow the A&A style

\newcommand{\Msol}{\ensuremath{M_{\odot}}}
\newcommand{\Lsol}{\ensuremath{L_{\odot}}}

\begin{document} 

%the mini-starburst ridge
   \title{Detection of a high-mass prestellar core candidate in W43-MM1\thanks{
        Figures~\ref{cs-spec} and \ref{siolines} are only available in electronic form at http://www.aanda.org.}}

  \author{T. Nony\inst{1}
          \and F. Louvet\inst{2} 
          \and F. Motte\inst{1,3}
          \and J. Molet\inst{4}
          \and K. Marsh\inst{5}
          \and E. Chapillon\inst{6,4} 
          \and A. Gusdorf\inst{7}
          \and N. Brouillet\inst{4}     
          \and S. Bontemps\inst{4}
          \and T. Csengeri\inst{8}
          \and D. Despois\inst{4} 
          \and Q. Nguyen Luong\inst{9}
          \and A. Duarte-Cabral\inst{5}
          \and A. Maury\inst{3}
          }

   \institute{Univ. Grenoble Alpes, CNRS, IPAG, 38000 Grenoble, France
   \and Departmento de Astronomia de Chile, Universidad de Chile, Santiago, Chile
   \and AIM, CEA, CNRS, Université Paris-Saclay, Université Paris Diderot, Sorbonne Paris Cité, 91191 Gif-sur-Yvette, France
   \and OASU/LAB, Univ. de Bordeaux - CNRS/INSU, 36615 Pessac, France
   \and School of Physics and Astronomy, Cardiff University, Queens Buildings, The Parade, Cardiff CF24 3AA, UK
   \and Institut de RadioAstronomie Millim\'etrique (IRAM), Grenoble, France
   \and LERMA, CNRS, Observatoire de Paris, \'Ecole Normale Sup\'erieure, 24 rue Lhomond, 75231 Paris Cedex 05, France 
   \and Max-Planck-Institut f\"ur Radioastronomie, Auf dem H\"ugel 69, 53121 Bonn, Germany
   \and Canadian Institute for Theoretical Astrophysics, University of Toronto, 60 St. George Street, Toronto, ON M5S 3H8, Canada   
                        }

   \date{}

% \abstract{}{}{}{}{} 
% 5 {} token are mandatory
 
  \abstract
   {}
     {To constrain the physical processes that lead to the birth of high-mass stars it is mandatory to study the very first stages of their formation.
     We search for high-mass analogs of low-mass prestellar cores in W43-MM1.}
   {We conducted a 1.3~mm ALMA mosaic of the complete W43-MM1 cloud, which has revealed numerous cores with $\sim\!2000$~au FWHM sizes. We investigated the nature of cores located at the tip of the main filament, where the clustering is minimum. We used the continuum emission to measure the core masses and the $^{13}$CS(5-4) line emission to estimate their turbulence level. We also investigated the prestellar or protostellar nature of these cores by searching for outflow signatures traced by CO(2-1) and SiO(5-4) line emission, and for molecular complexity typical of embedded hot cores.}
   {Two high-mass cores of $\sim$\,1300~ au diameter and $\sim$\,60$~\Msol$ mass are observed to be turbulent but gravitationally bound. One drives outflows and is associated with a hot core. The other core, W43-MM1\#6, does not yet reveal any star formation activity and thus is an excellent high-mass prestellar core candidate.} 
   {}
     
   \keywords{stars: formation – stars: protostars – stars: massive – ISM: clouds – submillimeter: ISM
               }
               
   \maketitle
%
%-------------------------------------------------------------------

\section{Introduction} 
Despite the large efforts made in the past ten years to improve our understanding of the formation of high-mass stars, two competing families of models remain. In the `core-fed' or `core-accretion' models, a high-mass star forms through the monolithic collapse of a massive, turbulent prestellar core that formed quasi-statically \citep[e.g.,][]{Mckee2003}. This preassembled core is in virial equilibrium, similar to low-mass prestellar cores, but supported by some magnetic and/or supersonic turbulent pressure that prevents the subfragmentation of this core. A few high-mass prestellar core candidates have been reported \citep{Bontemps2010,duarte2013co,Wang2014} but more examples are needed to support this scenario. 

On the other hand, the `clump-fed' models involve the gas mass reservoir surrounding individual cores through dynamical processes in their parental cloud. These models propose the rapid growth of cores via competitive accretion of the common cloud mass reservoir \citep[e.g.,][]{Bonnel06} or via accretion streams associated with the global hierarchical collapse of clouds \citep[e.g.,][]{smith09,vazquez17}. Observationally, there is a growing body of evidences in favor of dynamical, clump-fed models \citep[e.g.,][]{schneider10, csengeri11b, peretto13, henshaw14, louvet16}. In the empirical evolutionary sequence recently proposed for the formation of high-mass stars, the high-mass prestellar core phase does not even exist and high-mass protostellar cores form from low-mass protostellar cores, which accrete further material from their parental massive dense core (MDC) \citep{motte18a}. 

Imaging cold, high-mass star-forming clouds is needed to reach a consensus on the existence of a high-mass prestellar core phase. W43, located at 5.5\,kpc from the Sun \citep{zhang14}, contains two of the largest groups of molecular clouds in the first Galactic quadrant, among them W43-MM1 \citep{nguyen11b, nguyen13}. This $6\,$pc$^2$ ridge has a $2\times10^4\,M_\odot$ mass and qualifies as `mini-starburst' because its star formation activity is reminiscent of that of starburst galaxies 
\citep[SFR\,$\sim$\,6000\,$\Msol\,$Myr$^{-1}$;][]{motte03, louvet14}. Imaged with ALMA, W43-MM1 revealed one of the youngest and richest clusters of high-mass cores in the Milky Way \citep{motte18b}.

In the present paper, we characterize two high-mass cores located in the least clustered part of W43-MM1 and report on the possible discovery of one high-mass prestellar core. From the observations presented in Sect.~\ref{s:obs}, we derive the core characteristics using dust continuum (see Sect.~\ref{s:cont}) and investigate their gravitational boundedness and  evolutionary status with molecular line observations (see Sects.~\ref{s:13cs}--\ref{s:SF}). Core properties are then used in Sect.~\ref{s:disc} to discuss their prestellar versus protostellar nature and the most probable physical process forming high-mass stars in W43-MM1. 

\section{Observations and data reduction}
\label{s:obs}

Observations were carried out in Cycle 2 between July 2014 and June 2015 (project \#2013.1.01365.S), with ALMA 12~m and 7~m (ACA) arrays and baselines ranging from 7.6~m to 1045~m. W43-MM1 was imaged with a $78\arcsec\times$ 53$\arcsec$ (2.1 pc $\times$ 1.4 pc) mosaic composed of 33 fields with the 12~m array and 11 fields with ACA. In the 12~m configuration, the primary beam is of 26.7$\arcsec$ (45.8$\arcsec$ with ACA) and the maximum detectable scale is of 12$\arcsec$ (21$\arcsec$ with ACA). Parameters of the four spectral windows used in this study are presented in Table~\ref{spw}. Data were reduced in CASA 4.3.1, applying manual and self-calibration scripts. 

We then used the CLEAN algorithm with a robust weight of 0.5 and the multiscale option to minimize interferometric artifacts due to missing short spacings. Core extraction was carried out on the 12 m continuum data providing the best resolution (synthesized beam of 0.37$\arcsec$ $\times$ 0.53$\arcsec$, $\sim$2400 au at 5.5~kpc) and the best sensitivity (0.13 mJy\,beam$^{-1\,}$ on the 1.9~GHz averaged map). For the molecular line studies we used the merged (12~m + 7~m) data with characteristics given in Table~\ref{spw}. 
 
%--------------------------------------------------- One column table
\begin{table}[!hb]
   \caption[]{Parameters of the merged data spectral windows.}
   \label{spw}
    \begin{tabular}{ c c c c c c c }
    \hline
    \hline
    Spectral     &   $\nu_{\rm obs}$  & Bandwidth  & \multicolumn{2}{c}{Resolution}  & rms \\
    window           &  [GHz]        &   [MHz]    & \multicolumn{2}{c}{[$\arcsec$] ~ [km~s$^{-1}$] } & \tablefootmark{a} \\
     \hline
  
     SiO(5-4)    &  217.033  & 234  & 0.48 &  0.3 & 2.5 \\
     CO(2-1)    &  230.462   & 469  & 0.46 & 1.3 & 3.1 \\
     $^{13}$CS(5-4)  &  231.144  & 469 & 0.46 & 0.3  & 3.1 \\
    Continuum   & 233.4  &  1875 & 0.43  & 1.3  & 1.9 \\
    \hline
    \end{tabular}
    \tablefoottext{a}{$1\sigma$ rms in [mJy\,beam$^{-1}$]. 1 mJy\,beam$^{-1}$ corresponds to 0.12~K at 233.4 GHz.}
   \end{table}
%
%--------------------------------------------------------------------
 
\begin{figure}
%\vskip -1cm
\includegraphics[width=1.1\hsize]{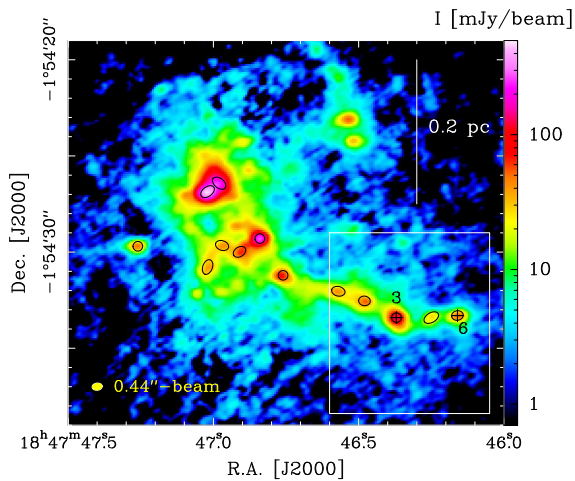}
%\vspace{-1.3cm}
\caption{Thirteen high-mass cores ($M_{\rm core} > 16\, \Msol$, black ellipses) discovered in the W43-MM1 protocluster \citep[see][]{motte18b}. The white box outlines the zoom of Figs.~\ref{co-s6}a--b toward two very massive cores in the least clustered part of the main filament (\#3 and \#6, black crosses). }
\label{cont}
\end{figure}
   
\begin{table*}
   \caption[]{Cores' main characteristics derived from ALMA observations. }
   \label{cores}
    \begin{tabular}{ c  c  c  c  c  c  c  c  c c c}
    \hline
    \hline
     Cores    & R.A. & Dec. &  FWHM   & $S^{\rm peak}_{\rm1.3 mm}$ & $S^{\rm int}_{\rm 1.3 mm}$ & $T_{\rm dust}$ & $M_{\rm core}$  & $\Delta V_{\rm^{13}CS}$ & $\alpha_{\rm vir}$\tablefootmark{a} & Outflow lobes\\
           & [J2000]  & [J2000]   &  [au]           &    [mJy/beam]               & [mJy]        &   [K]       & [$\Msol$] & [km~s$^{-1}$]  & & [pc]  \\
     \hline
      \#3   & 18:47:46.37  & -1:54:33.41  &  1200  & $109 \pm 2$ & $222 \pm 2$  & $45 \pm 2$ & $59 \pm 4$ & $4.0 \pm 0.3$  & 0.2  & 0.12 \\
     \#6   &  18:47:46.16 & -1:54:33.30  &  1300  & $46.8 \pm 0.8$  & $94 \pm 1$  & $23 \pm 2$  &  $56 \pm 9$ & $3.3 \pm 0.2$  & 0.2-0.3 & <\,0.01  \\
    \hline
    \end{tabular}
    \\
    \tablefootmark{a}{Virial parameter, $\alpha_{\rm vir}=M_{\rm vir}/M_{\rm core}$, calculated with density indexes of $n=2$  for both cores and $n=0$ for core \#6.~}
   \end{table*}

\section{Result and analysis}
\label{s:res}

\subsection{Core extraction and mass estimation } 
\label{s:cont}
Compact sources were extracted using getsources (v1.140127), a multiscale, multiwavelength source-extraction algorithm \citep{men2012multi}. The final getsources catalogs of the W43-MM1 cloud contains 131 reliable sources presented in \cite{motte18b}. 
Among these sources, 13 high-mass cores with masses from 16 to 100 $\Msol$ are expected to form high-mass stars, assuming a 50\% core-to-star efficiency for these very high-density cores. They have diameters ranging from 1200 to 2600~au once deconvolved from the 0.44$\arcsec$ beam. These cores are outlined with ellipses on the 1.3 mm continuum map of Fig~\ref{cont}. We chose to focus on cores \#3 and \#6, two high-mass cores located at the western tip of the main filament (see Fig.~\ref{cont}), where the blending associated with clustering is minimum. Table~\ref{cores} lists their coordinates, Gaussian full width at half maximum (FWHM) sizes, peak, and integrated fluxes.

The (gas + dust) mass of a core, having uniform opacity throughout its solid angle, is given by
        \begin{equation}
        M_{\rm core} = - \dfrac{\Omega_{\rm b }  \, d^{2} }{\kappa_{\rm 1.3 mm}}\,
        \ln\left(1\,-\,\frac{S^{\rm peak}_{\rm1.3 mm}}{\Omega_{\rm b}\;B_{1.3{\rm mm}}(T_{\rm dust})}\right) \times \dfrac{S^{\rm int}_{\rm 1.3 mm}}{S^{\rm peak}_{\rm 1.3 mm}}, \\ 
        \label{Mcore}
        \end{equation}
        
where $\Omega_{\rm b}$ is the solid angle of the beam, $d\,=\,5.5$ kpc is the distance to the core, $\kappa_{\rm 1.3 mm}$ is the dust opacity per unit of (gas $+$ dust) mass at 1.3 mm, $S^{\rm int}_{\rm 1.3 mm}$ and $S^{\rm peak}_{\rm 1.3 mm}$ are the peak and integrated fluxes of the core and $B_{1.3{\rm mm}}(T_{\rm dust})$ is the Planck function at dust temperature $T_{\rm dust}$. 

The 1.3 mm continuum flux of a core arises mainly from thermal dust emission, which is generally optically thin. This is no longer the case for the densest, most massive cores. Equation~\ref{Mcore} thus includes a correction for dust opacity, which increases the mass of cores \#3 and \#6 by 20\%. Fluxes given in Table~\ref{cores} are also corrected for contamination by free-free and line emission \citep[see details in][]{motte18b}. 
The mean dust temperature of cores was estimated by applying the Bayesian PPMAP procedure \citep{marsh2015} to all existing continuum data of W43-MM1 \citep[see Fig.~3 of][]{motte18b}.
We adopted $\kappa_{\rm 1.3 mm} = {\rm 0.01~ cm^{2}g^{-1}}$ as being the most appropriate for the high-density, $n_{\rm H_2} \sim 10^9 \rm cm^{-3} $, cool to warm, $20-40~K$, cores. The absolute uncertainties on core masses is estimated to be about a factor of two. Table~\ref{cores} lists the assumed temperatures and derived masses, $M_{\rm core}=59\,\Msol$ and $56\,\Msol$ for cores \#3 and \#6, respectively.

\subsection{Virial mass and gravitational boundedness}
\label{s:13cs}

To qualify as stellar progenitors, cores must be gravitationally bound. We characterize the isotropic turbulence of cores listed in Table~\ref{cores} using the J=5-4 transition of $^{13}$CS, a good tracer of high-density gas (critical density $\sim\!10^6 \rm cm^{-3}$). The $^{13}$CS emission peaks toward high-mass cores but is also detected along their parental filament and their outflow. We thus subtracted, from each core spectrum, the contribution of the core parental filament and outflow before fitting a Gaussian (see Fig.~\ref{cs-spec}). 
The line FWHM given in Table~\ref{cores} corresponds to the average of the fits before and after this subtraction. The line widths are large, $\Delta V_{\rm ^{13}CS} \sim 4.0~\rm km\,s^{-1}$ and $\sim\!3.3~\rm km\,s^{-1}$, for cores \#3 and \#6, respectively. 

The virial mass, $M_{\rm vir}$, is measured as 
\begin{equation}
M_{\rm vir} = 3 \left( \dfrac{5-2n}{3-n} \right) \dfrac{\sigma_{\rm NT} ^2 \, \rm FWHM}{G},
\label{Mvir}
\end{equation}

where n is the index of the density profile ($\rho \propto r^{-n}$), $\sigma _{\rm NT}$ is the 1D nonthermal velocity dispersion, FWHM is the deconvolved size of cores, and G is the gravitational constant. We assume a density index of $n=2$ (case of a centrally concentrated protostar) for both cores and $n=0$ (case of a flat starless core) for core \#6, whose nature could be prestellar (see Sect.~\ref{s:disc}). The parameter $\sigma _{\rm NT}$ is derived by subtracting the thermal component from the total velocity dispersion, 
$\sigma _{\rm NT}^2 = \sigma^2 - \sigma_{\rm th}^2 = \frac{ \Delta V_{\rm ^{13}CS}}{8 \rm ln(2)}^2 - \frac{k_{\rm B} \, T_{\rm dust}}{\mu_{\rm ^{13}CS} \, m_{\rm H}},  $
 where $\mu_{\rm ^{13}CS}=45$ is the $^{13}$CS molecular weight, $k_{\rm B}$ is the Boltzmann constant, and $m_{\rm H}$ is the hydrogen mass. Given that their virial parameters are well below unity, $\alpha_{\rm vir}\,=\,M_{\rm vir}/M_{\rm core}=0.2-0.3$, even with a factor of two uncertainty on masses, cores \#3 and \#6 are gravitationally bound and could be collapsing if no extra support, such as from magnetic fields, prevents it.

%-----------------------------------------------------------------
%                                                One column figure

   \begin{figure}
   %\vskip -0.8cm
  \centerline{ \hskip 2cm \includegraphics[width=1.05\hsize]{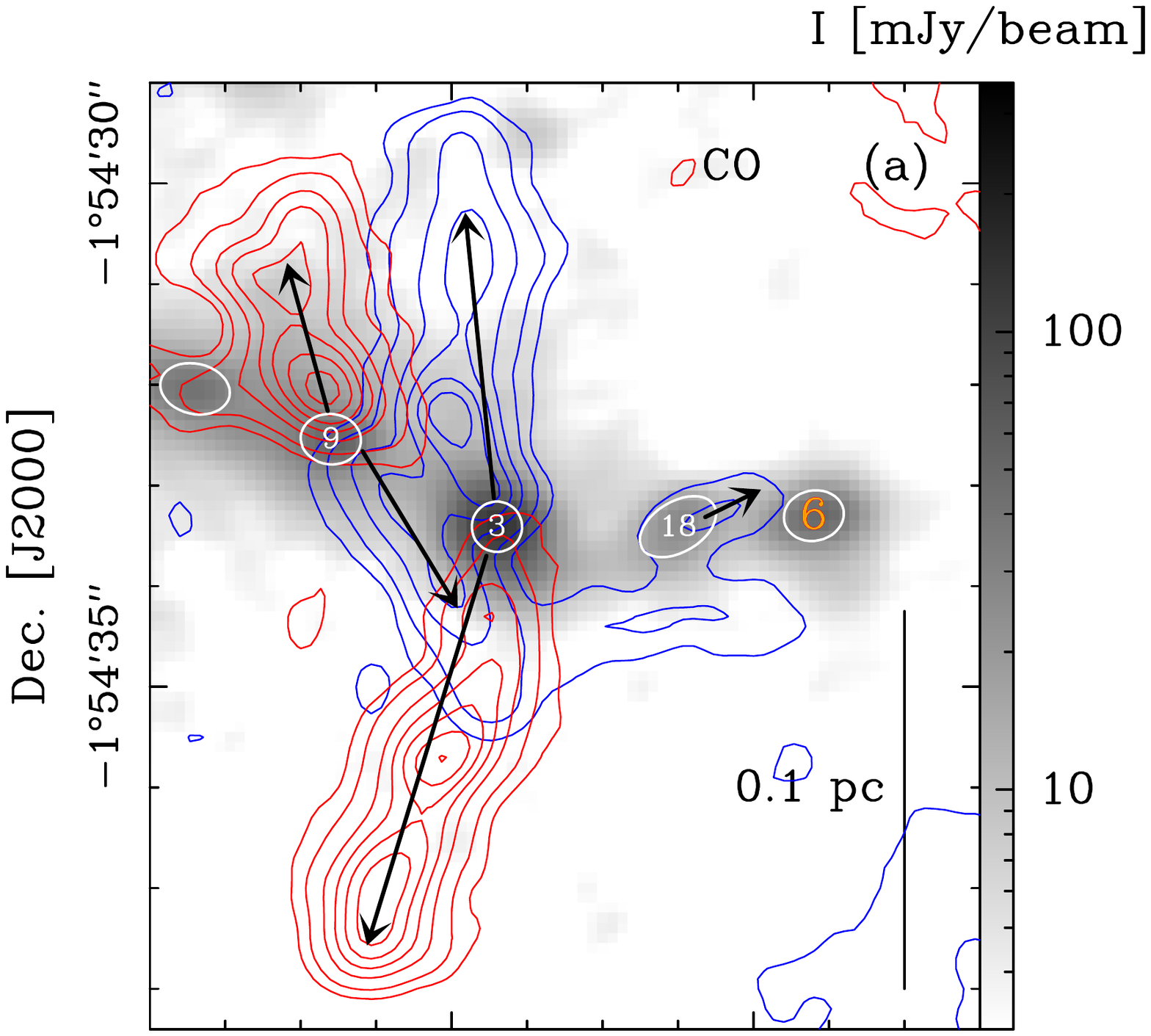}}
   %\vskip -2.3cm
   \centerline{ \hskip 2cm \includegraphics[width=1.05\linewidth]{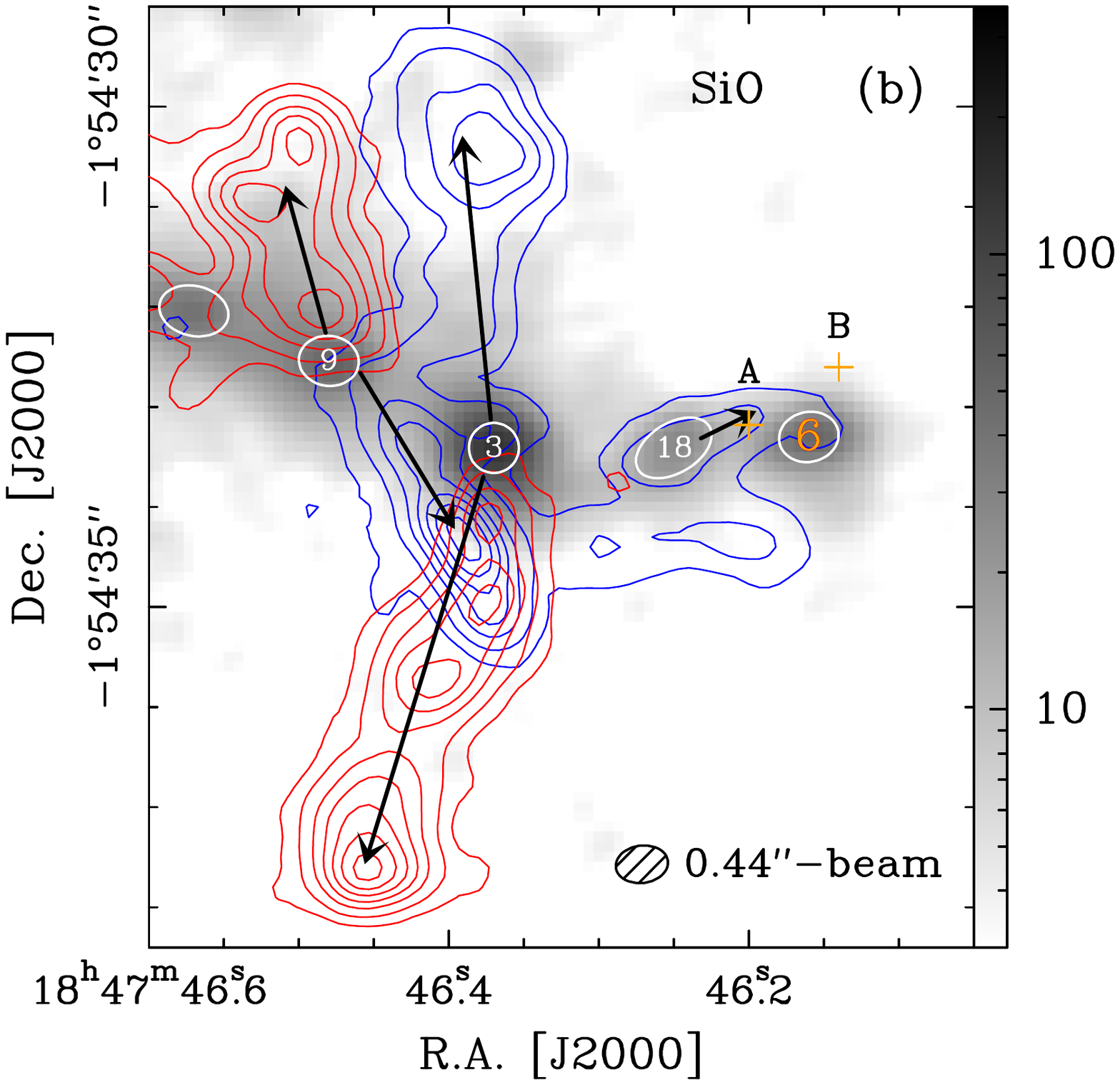}}
   %\vskip -0.5cm
      \caption{Outflows driven by cores \#3, \#9, and \#18. Blue and red contours from the CO(2-1) \textbf{(a)} and SiO(5-4) \textbf{(b)} line wings are overplotted on the 1.3 mm continuum emission in grayscale. Lines are integrated over $38-88\,\rm km\,s^{-1}$ ($60-90\,\rm km\,s^{-1}$) for the CO (resp. SiO) blue contours, and over $108-158\,\rm km\,s^{-1}$ ($104-134\,\rm km\,s^{-1}$) for the CO (resp. SiO) red contours. CO contour levels are, in units of $\sigma_{\rm CO}=86\,\rm mJy\,beam^{-1}\,km\,s^{-1}$, 6 and 15 to 115 by 20 steps. They are for SiO, in unit of $\sigma_{\rm SiO}=18.5\,\rm mJy\,beam^{-1}\,km\,s^{-1}$, 5 and from 15 to 90 by 15 steps. Ellipses represent the FWHM diameter of cores extracted by getsources, crosses indicate locations A and B, and arrows indicate the outflow directions.  
              }
         \label{co-s6}
   \end{figure}
   
\subsection{Search for signposts of protostellar activity}
\label{s:SF}

Gas outflows associated with accretion are certainly the best, systematic signpost of protostellar activity.
We integrated the high-velocity line wings of CO(2-1) and SiO(5-4) around the systemic velocities $V_{\rm LSR} \simeq 96~\rm km\,s^{-1}$ for core \#6 and $98~ \rm km\,s^{-1}$ for core \#3. The resulting lobes of bipolar outflows shown in Figs.~\ref{co-s6}a--b present an excellent agreement between the CO(2-1) and SiO(5-4) line tracers. While core \#3 unambiguously drives a well-developed outflow (see Figs.~\ref{co-s6}a--b and Table~\ref{cores}), the weak blue-shifted emission detected both in CO(2-1) and SiO(5-4) toward core \#6 is probably not associated with this core. As shown in Fig.~\ref{siolines}, the SiO(5-4) emission consists of two components. One decreasing from core \#18 to location~A (see Fig.~\ref{co-s6}b) is associated with the outflow of core \#18. Another component peaking at location~B could be associated with low-velocity shocks such as those observed by \cite{louvet16}. 
Core \#6 thus seems to lack protostellar outflow but more sensitive and higher resolution observations (see Table~\ref{cont}) will be needed to confirm this finding.
 
  \begin{figure}[!h]
   \centerline{\includegraphics[width=1.1\linewidth]{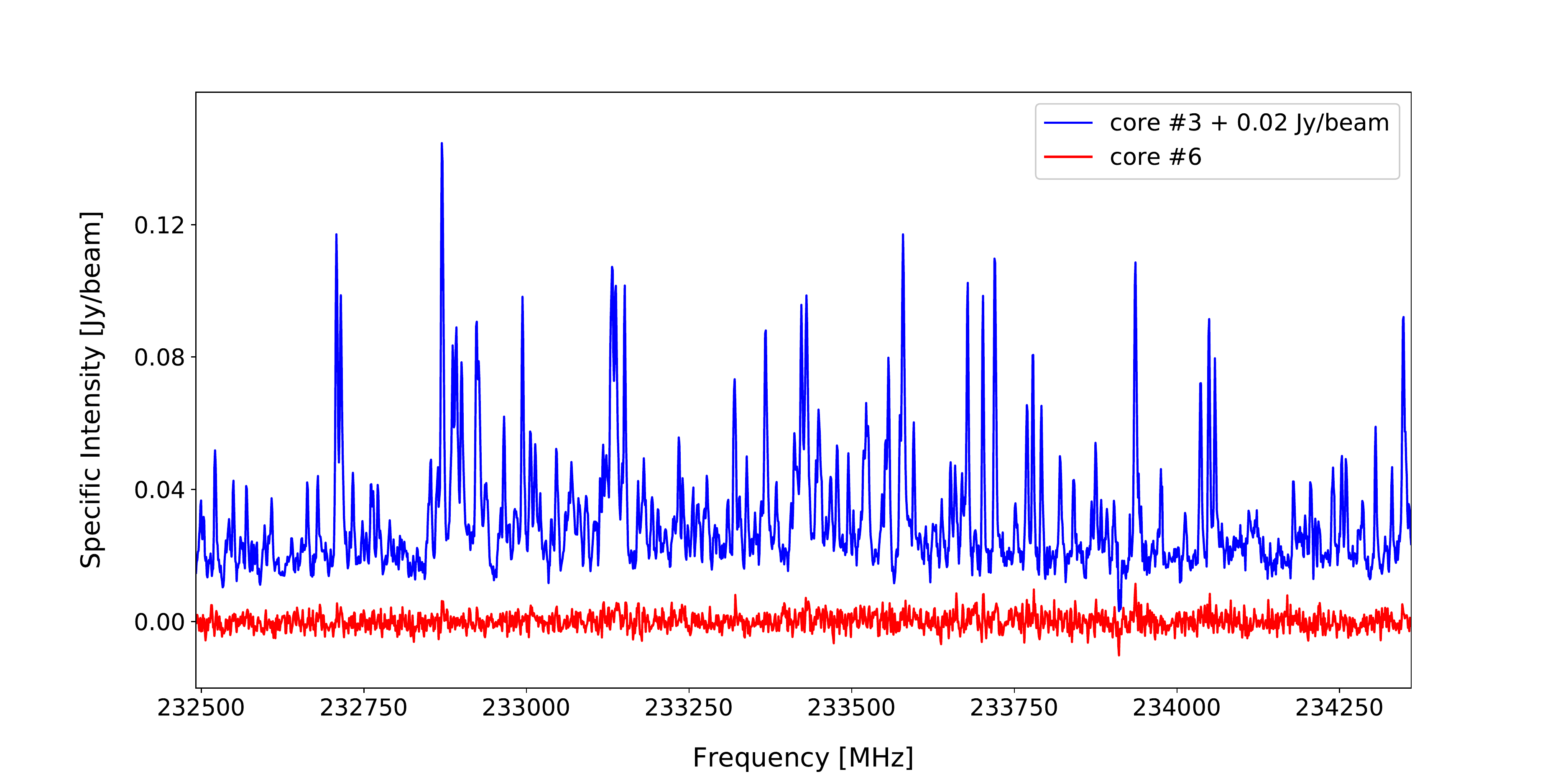}}
      \caption{Continuum-subtracted spectra of the 1.9~GHz band used to estimate the 1.3~mm continuum emission of cores. Core \#3 stands out with a rich line forest, whereas very few lines are observed toward core \#6. }
         \label{cont-spec}
   \end{figure}
   
The detection of line forests in spectral bands indicates molecular complexity \citep[e.g.,][]{Herbst09} and is often used to pinpoint hot cores. Figure~\ref{cont-spec} shows the 1.9 GHz spectral band at 233.5 GHz for cores \#3 and \#6. Very few lines are detected toward core \#6: few transition lines associated with methanol and methyl formate are detected at a $\sim$3$\sigma$ level in the continuum spectral window of core \#6. A careful analysis and modeling of the peculiar chemical complexity toward this core (Molet et al. in prep.) aims to interpret it in terms of protostellar heating, small-scale, or even large-scale shocks. These features are in strong contrast with the line forest observed for core \#3 (see Fig.~\ref{cont-spec}).  
We also performed a quantitative measurement of the richness of the line forest by comparing the flux integrated over a line-free composite band of 65 MHz with the flux integrated over the 1.9 GHz full band, consisting of the sum of the line-free plus line-contamination fluxes. Line contamination, estimated by its the ratio to the line-free continuum flux, is $>\!13\%$ for core \#3 and $\sim\!0\%$ toward core \#6 \citep{motte18b}. 
These results suggest that core \#3 hosts a hot core while core \#6 may not yet.
 
\section{Discussion and conclusions}
\label{s:disc}
\begin{figure}
\centerline{ \includegraphics[width=1.0\linewidth]{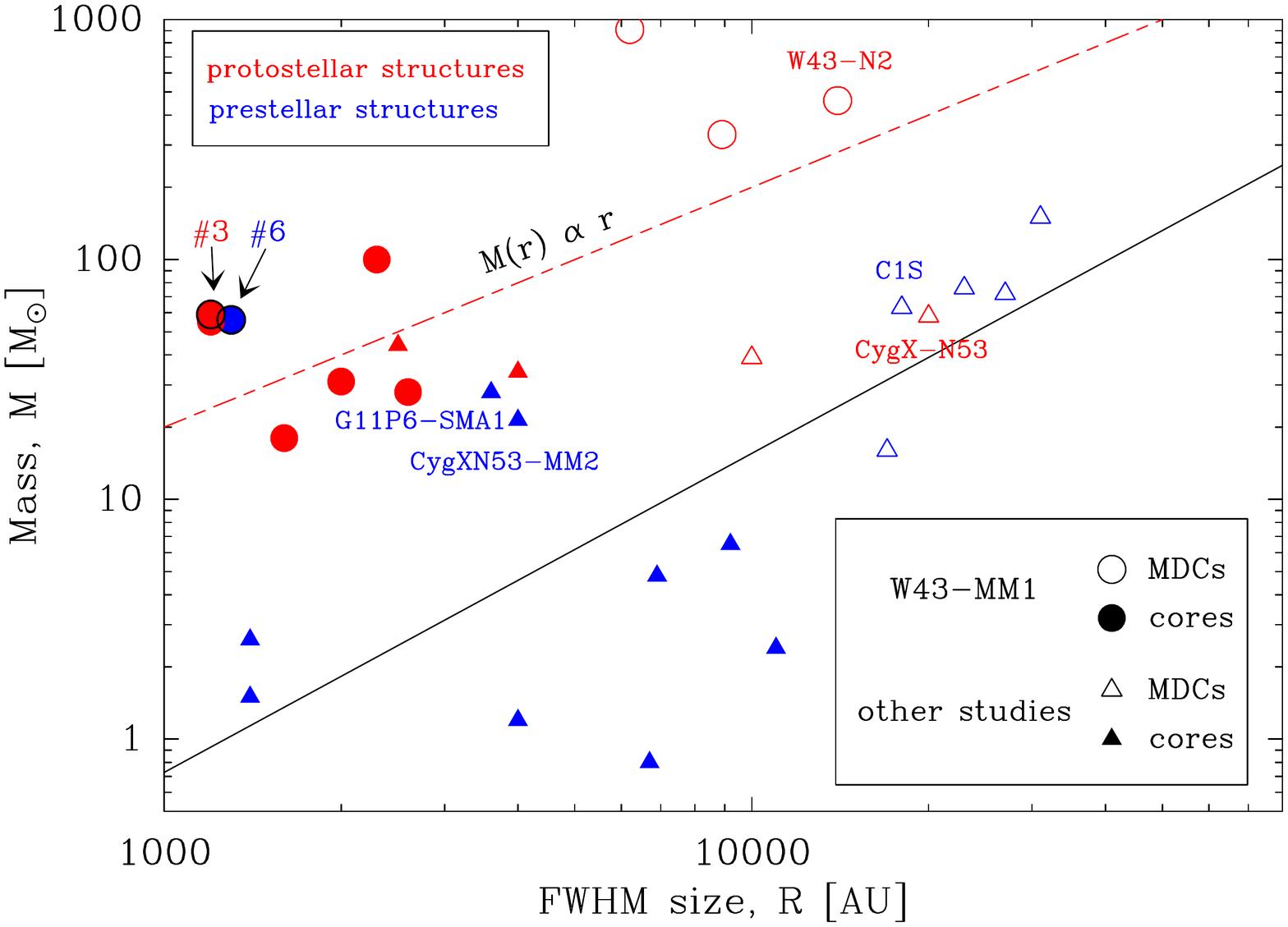} }
%\vskip -0.7cm
\caption{Extremely massive $\sim$1000~au-scale cores \#3 and \#6 and their parental MDCs in W43-MM1 \citep[circles;][]{louvet14, motte18b}, compared to cores and MDCs of references studies \citep[triangles;][Louvet subm.]{Bontemps2010, Tan13, Wang2014, Tige17}. In these studies, masses are calculated from millimeter continuum with the the same assumptions: dust temperature evaluated from large-scale spectral energy distribution fit (except for \citealt{Wang2014} which use $\rm NH_3$), gas-to-dust ratio of 100 (except for \citealt{Tan13} which use 141), and an opacity index of 1.5. Cores \#3 and \#6 lie well above the empirical massive star formation threshold proposed by \cite{kauffmann10} (black line).
}
\label{M-R}
\end{figure}

The two cores discussed in this work have remarkably large masses, $\sim$\,$60~\Msol$, given their small diameters, $\sim$1300~au (see Table~\ref{cores}), leading to extreme densities, $n_{\rm H_2} \sim 10^9\, \rm cm^{-3} $. In Fig.~\ref{M-R}, cores \#3 and \#6 are compared to cores of a few 1000~au sizes found within cold MDCs (0.1~pc). 
These two cores are among the most massive detected so far at $1000-2000$~au scales. They have high turbulent levels, $\sigma \sim 1.5~\rm km\,s^{-1}$ corresponding to Mach number $\sim 4.5$, which resemble those observed for ten-times larger young MDCs \citep[e.g.,][]{ragan12b, kauffmann13b} and probably testifies to the high dynamics observed in W43-MM1 \citep[e.g.,][]{nguyen13, louvet16}. Despite that, these cores have low virial parameters,  $\alpha_{\rm vir}\,=0.2-0.3$, which proves they are gravitationally bound and could be collapsing if no extra support such as magnetic field prevents it. Core \#3 drives a bipolar outflow (see Figs.~\ref{co-s6}a--b) and displays molecular complexity possibly associated with hot core emission (see Fig.~\ref{cont-spec}).
In contrast, core \#6 shows no signs of protostellar activity. We currently cannot rule out the existence of an unresolved outflow and/or a small and embedded hot core, but core \#6 is, for now, an excellent high-mass prestellar core candidate.

Core \#6 could therefore be among the very few examples representing the initial conditions of the \cite{Mckee2003} model of high-mass star formation. 
In detail, this turbulent-core accretion model assumes that a $\sim\!12\,000~ \rm au$ core with $\sim\!60~\Msol$ and $\sigma \sim 1~\rm km\,s^{-1}$ is quasi-statically assembled. Observationally, however, most of the $10\,000~ \rm au$ cloud structures subfragment in protostars \citep{Bontemps2010, Tan16} or dissolve into low-mass cores \citep[][Louvet et al. subm.]{Kong17}.
In contrast core \#6, with its $56~\Msol$ mass within 1300~au, keeps a large fraction, $\sim$12\%, of its parental MDC mass \citep[W43-N2,][]{louvet14}. This gas concentration tends to follow the $M(r) \propto r$ relation predicted for gravitationally bound cores such as Bonnor Ebert spheres (see Fig.~\ref{M-R}).
Core \#6 is also twice more massive than the few other high-mass prestellar core candidates, including CygXN53-MM2 \citep{Bontemps2010, duarte2013co} and G11P6-SMA1 \citep{Wang2014}. 

Given that cores \#3 and \#6 have the same masses but different evolutionary stages, we use these cores to illustrate the evolutionary sequence proposed by \cite{motte18a}. Core \#3 is a high-mass core, hosting a stellar embryo massive enough to power a hot core. This would then correspond to the IR-quiet or IR-bright protostellar core phases, which are steps~4 and 5 of this scenario. Core \#6 itself could either be a high-mass prestellar core or a very young high-mass protostellar core, hosting a very low-mass stellar embryo. While the existence of a high-mass prestellar phase is refuted in \cite{motte18a}, the second interpretation would put core \#6 in the IR-quiet protostellar core phase (step~4).

The mass-to-luminosity ratio, often used as an indicator of the evolutionary stage of cores, 
complies with core \#6 being younger than core \#3.
\cite{motte18b} roughly evaluated the bolometric luminosity of cores from the total luminosity of W43-MM1, $\sim\,2 \times 10^4\, \Lsol$ \citep{motte03}, divided between the cores in proportion to their line contamination. These luminosities, $L \sim 10~\Lsol$ for cores \#6 and $\sim\!1000~\Lsol$ for cores \#3, lead to ratios of $M_{\rm core}/L_{\rm bol} \sim 6\, \Msol/\Lsol$ and $\sim\!0.06\,\Msol/\Lsol$ for cores \#6 and \#3, respectively. 
These decreasing ratios, for cores of similar sizes and masses, indicates an increasing proportion of gas in the stellar embryo relative to the cold core and thus an evolution in time through the high-mass star-formation process.

We have proposed that core \#6 could represent the initial conditions of the turbulent-core monolithic-collapse model \citep{Mckee2003}. However, since W43-MM1 is a very dynamical cloud \citep{louvet16}, its cores are expected to simultaneously collapse and grow in mass from their surrounding cloud gas. Therefore, core \#6 may be more consistent with the earliest stage of the accretion-stream model of \cite{smith09} and \cite{vazquez17} than with the quasi-static preassembled core taken for initial conditions in the model of \cite{Mckee2003}.

\begin{acknowledgements}
We thank A. Men'shchikov for his help on getsources. 
This paper makes use of the following ALMA data: \#2013.1.01365.S. ALMA is a partnership of ESO (representing its member states), NSF (USA), and NINS (Japan), together with NRC (Canada), MOST and ASIAA (Taiwan), and KASI (Republic of Korea), in cooperation with the Republic of Chile. The Joint ALMA Observatory is operated by ESO, AUI/NRAO, and NAOJ.
This project has received funding from the European Union's Horizon 2020 research and innovation program StarFormMapper under grant agreement No 687528.
This work was supported by the Programme National de Physique Stellaire and Physique et Chimie du Milieu Interstellaire (PNPS and PCMI) of CNRS/INSU (with INC/INP/IN2P3) co-funded by CEA and CNES.
\end{acknowledgements}


\bibliographystyle{aa}
\bibliography{biblio-letter}    

\begin{thebibliography}{29}
\expandafter\ifx\csname natexlab\endcsname\relax\def\natexlab#1{#1}\fi

\bibitem[{{Bonnell} \& {Bate}(2006)}]{Bonnel06}
{Bonnell}, I.~A. \& {Bate}, M.~R. 2006, \mnras, 370, 488

\bibitem[{{Bontemps} {et~al.}(2010){Bontemps}, {Motte}, {Csengeri}, \&
  {Schneider}}]{Bontemps2010}
{Bontemps}, S., {Motte}, F., {Csengeri}, T., \& {Schneider}, N. 2010, \aap, 524

\bibitem[{{Csengeri} {et~al.}(2011){Csengeri}, {Bontemps}, {Schneider},
  {Motte}, \& {Dib}}]{csengeri11b}
{Csengeri}, T., {Bontemps}, S., {Schneider}, N., {Motte}, F., \& {Dib}, S.
  2011, \aap, 527, A135

\bibitem[{Duarte-Cabral {et~al.}(2013)Duarte-Cabral, Bontemps, Motte,
  Hennemann, Schneider, \& Andr{\'e}}]{duarte2013co}
Duarte-Cabral, A., Bontemps, S., Motte, F., {et~al.} 2013, \aap, 558, A125

\bibitem[{{Henshaw} {et~al.}(2014){Henshaw}, {Caselli}, {Fontani},
  {Jim{\'e}nez-Serra}, \& {Tan}}]{henshaw14}
{Henshaw}, J.~D., {Caselli}, P., {Fontani}, F., {Jim{\'e}nez-Serra}, I., \&
  {Tan}, J.~C. 2014, \mnras, 440, 2860

\bibitem[{{Herbst} \& {van Dishoeck}(2009)}]{Herbst09}
{Herbst}, E. \& {van Dishoeck}, E.~F. 2009, \araa, 47, 427

\bibitem[{{Kauffmann} \& {Pillai}(2010)}]{kauffmann10}
{Kauffmann}, J. \& {Pillai}, T. 2010, \apj, 723, L7

\bibitem[{{Kauffmann} {et~al.}(2013){Kauffmann}, {Pillai}, \&
  {Goldsmith}}]{kauffmann13b}
{Kauffmann}, J., {Pillai}, T., \& {Goldsmith}, P.~F. 2013, \apj, 779, 185

\bibitem[{{Kong} {et~al.}(2017){Kong}, {Tan}, {Caselli}, {Fontani}, {Liu}, \&
  {Butler}}]{Kong17}
{Kong}, S., {Tan}, J.~C., {Caselli}, P., {et~al.} 2017, \apj, 834, 193

\bibitem[{{Louvet} {et~al.}(2016){Louvet}, {Motte}, {Gusdorf}, {Nguy{\^e}n
  Luong}, {Lesaffre}, {Duarte-Cabral}, {Maury}, {Schneider}, {Hill}, {Schilke},
  \& {Gueth}}]{louvet16}
{Louvet}, F., {Motte}, F., {Gusdorf}, A., {et~al.} 2016, \aap, 595, A122

\bibitem[{{Louvet} {et~al.}(2014){Louvet}, {Motte}, {Hennebelle}, {Maury},
  {Bonnell}, {Bontemps}, {Gusdorf}, {Hill}, {Gueth}, {Peretto},
  {Duarte-Cabral}, {Stephan}, {Schilke}, {Csengeri}, {Nguyen Luong}, \&
  {Lis}}]{louvet14}
{Louvet}, F., {Motte}, F., {Hennebelle}, P., {et~al.} 2014, \aap, 570, A15

\bibitem[{Marsh {et~al.}(2015)Marsh, Whitworth, \& Lomax}]{marsh2015}
Marsh, K., Whitworth, A., \& Lomax, O. 2015, \mnras, 454, 4282

\bibitem[{{McKee} \& {Tan}(2003)}]{Mckee2003}
{McKee}, C.~F. \& {Tan}, J.~C. 2003, \apj, 585, 850

\bibitem[{Men’shchikov {et~al.}(2012)Men’shchikov, Andr{\'e}, Didelon,
  Motte, Hennemann, \& Schneider}]{men2012multi}
Men’shchikov, A., Andr{\'e}, P., Didelon, P., {et~al.} 2012, \aap, 542, A81

\bibitem[{{Motte} {et~al.}(2018{\natexlab{a}}){Motte}, {Bontemps}, \&
  {Louvet}}]{motte18a}
{Motte}, F., {Bontemps}, S., \& {Louvet}, F. 2018{\natexlab{a}}, \araa, 56
  [\eprint{https://doi.org/10.1146/annurev-astro-091916-055235}]

\bibitem[{{Motte} {et~al.}(2018{\natexlab{b}}){Motte}, {Nony}, {Louvet},
  {Marsh}, {Bontemps}, {Whitworth}, {Men'shchikov}, {Nguyen Luong}, {Csengeri},
  {Maury}, {Gusdorf}, {Chapillon}, {K{\"o}nyves}, {Schilke}, {Duarte-Cabral},
  {Didelon}, \& {Gaudel}}]{motte18b}
{Motte}, F., {Nony}, T., {Louvet}, F., {et~al.} 2018{\natexlab{b}}, Nature
  Astronomy, 2, 478

\bibitem[{{Motte} {et~al.}(2003){Motte}, {Schilke}, \& {Lis}}]{motte03}
{Motte}, F., {Schilke}, P., \& {Lis}, D.~C. 2003, \apj, 582, 277

\bibitem[{{Nguyen Luong} {et~al.}(2013){Nguyen Luong}, {Motte}, {Carlhoff},
  {Louvet}, {Lesaffre}, {Schilke}, {Hill}, {Hennemann}, {Gusdorf}, {Didelon},
  {Schneider}, {Bontemps}, {Duarte-Cabral}, {Menten}, {Martin}, {Wyrowski},
  {Bendo}, {Roussel}, {Bernard}, {Bronfman}, {Henning}, {Kramer}, \&
  {Heitsch}}]{nguyen13}
{Nguyen Luong}, Q., {Motte}, F., {Carlhoff}, P., {et~al.} 2013, \apj, 775, 88

\bibitem[{{Nguyen Luong} {et~al.}(2011){Nguyen Luong}, {Motte}, {Schuller},
  {Schneider}, {Bontemps}, {Schilke}, {Menten}, {Heitsch}, {Wyrowski},
  {Carlhoff}, {Bronfman}, \& {Henning}}]{nguyen11b}
{Nguyen Luong}, Q., {Motte}, F., {Schuller}, F., {et~al.} 2011, \aap, 529, A41

\bibitem[{{Peretto} {et~al.}(2013){Peretto}, {Fuller}, {Duarte-Cabral},
  {Avison}, {Hennebelle}, {Pineda}, {Andr{\'e}}, {Bontemps}, {Motte},
  {Schneider}, \& {Molinari}}]{peretto13}
{Peretto}, N., {Fuller}, G.~A., {Duarte-Cabral}, A., {et~al.} 2013, \aap, 555,
  A112

\bibitem[{{Ragan} {et~al.}(2012){Ragan}, {Heitsch}, {Bergin}, \&
  {Wilner}}]{ragan12b}
{Ragan}, S.~E., {Heitsch}, F., {Bergin}, E.~A., \& {Wilner}, D. 2012, \apj,
  746, 174

\bibitem[{{Schneider} {et~al.}(2010){Schneider}, {Csengeri}, {Bontemps},
  {Motte}, {Simon}, {Hennebelle}, {Federrath}, \& {Klessen}}]{schneider10}
{Schneider}, N., {Csengeri}, T., {Bontemps}, S., {et~al.} 2010, \aap, 520, A49+

\bibitem[{{Smith} {et~al.}(2009){Smith}, {Longmore}, \& {Bonnell}}]{smith09}
{Smith}, R.~J., {Longmore}, S., \& {Bonnell}, I. 2009, \mnras, 400, 1775

\bibitem[{{Tan} {et~al.}(2013){Tan}, {Kong}, {Butler}, {Caselli}, \&
  {Fontani}}]{Tan13}
{Tan}, J.~C., {Kong}, S., {Butler}, M.~J., {Caselli}, P., \& {Fontani}, F.
  2013, \apj, 779, 96

\bibitem[{{Tan} {et~al.}(2016){Tan}, {Kong}, {Zhang}, {Fontani}, {Caselli}, \&
  {Butler}}]{Tan16}
{Tan}, J.~C., {Kong}, S., {Zhang}, Y., {et~al.} 2016, \apj, 821, L3

\bibitem[{{Tig{\'e}} {et~al.}(2017){Tig{\'e}}, {Motte}, {Russeil}, {Zavagno},
  {Hennemann}, {Schneider}, {Hill}, {Nguyen Luong}, {Di Francesco}, {Bontemps},
  {Louvet}, {Didelon}, {K{\"o}nyves}, {Andr{\'e}}, {Leuleu}, {Bardagi},
  {Anderson}, {Arzoumanian}, {Benedettini}, {Bernard}, {Elia}, {Figueira},
  {Kirk}, {Martin}, {Minier}, {Molinari}, {Nony}, {Persi}, {Pezzuto},
  {Polychroni}, {Rayner}, {Rivera-Ingraham}, {Roussel}, {Rygl}, {Spinoglio}, \&
  {White}}]{Tige17}
{Tig{\'e}}, J., {Motte}, F., {Russeil}, D., {et~al.} 2017, \aap, 602, A77

\bibitem[{{V{\'a}zquez-Semadeni} {et~al.}(2017){V{\'a}zquez-Semadeni},
  {Gonz{\'a}lez-Samaniego}, \& {Col{\'{\i}}n}}]{vazquez17}
{V{\'a}zquez-Semadeni}, E., {Gonz{\'a}lez-Samaniego}, A., \& {Col{\'{\i}}n}, P.
  2017, \mnras, 467, 1313

\bibitem[{{Wang} {et~al.}(2014){Wang}, {Zhang}, {Testi}, {van der Tak}, {Wu},
  {Zhang}, {Pillai}, {Wyrowski}, {Carey}, {Ragan}, \& {Henning}}]{Wang2014}
{Wang}, K., {Zhang}, Q., {Testi}, L., {et~al.} 2014, \mnras, 439, 3275

\bibitem[{{Zhang} {et~al.}(2014){Zhang}, {Moscadelli}, {Sato}, {Reid},
  {Menten}, {Zheng}, {Brunthaler}, {Dame}, {Xu}, \& {Immer}}]{zhang14}
{Zhang}, B., {Moscadelli}, L., {Sato}, M., {et~al.} 2014, \apj, 781, 89

\end{thebibliography}
   
\begin{appendix}

\renewcommand{\thefigure}{A\arabic{figure}}

\setcounter{figure}{0}
%\vskip -2cm 
\begin{figure*}[!h]
   \centering
   %\vspace{-3cm}
   \includegraphics[width=1.0\hsize]{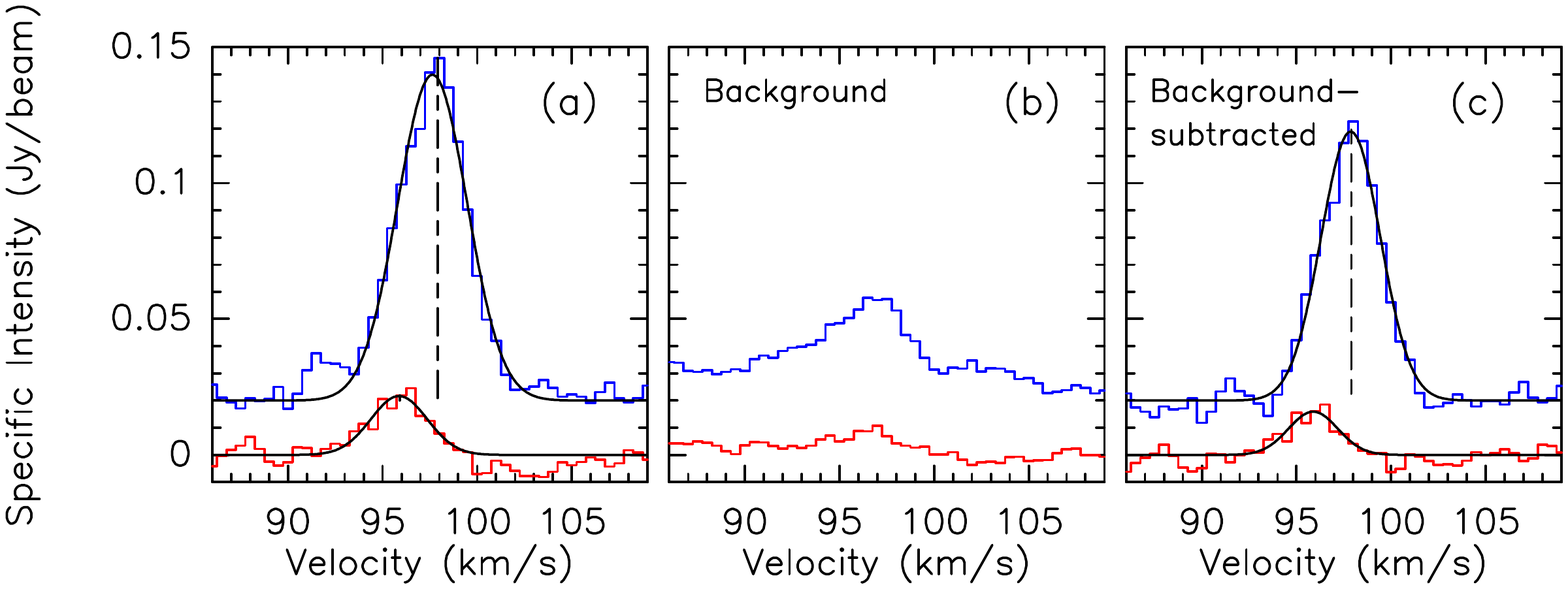}
 
    %\vspace{-3.5cm}
      \caption{Spectra of the $^{13}$CS line detected toward the continuum peak of core \#3 (in blue with a 0.02 Jy/beam offset) and core \#6 (in red), before (\textbf{a}) and after (\textbf{c}) subtraction of their parental filament and outflow emission (\textbf{b}). Black curves are the Gaussian fits used to measure the  line widths of the cores: $\Delta V_{\rm ^{13}CS}\,= 4.3~\rm km\,s^{-1}$ and $3.7~\rm km\,s^{-1}$ for core \#3, $\Delta V_{\rm ^{13}CS}\,= 3.5~\rm km\,s^{-1}$ and $3.1~\rm km\,s^{-1}$ for core \#6 before and after background subtraction, respectively.}
         \label{cs-spec}
   \end{figure*}

%\vskip -2cm 

\begin{figure*}[!h]
\includegraphics[width=1.05\hsize]{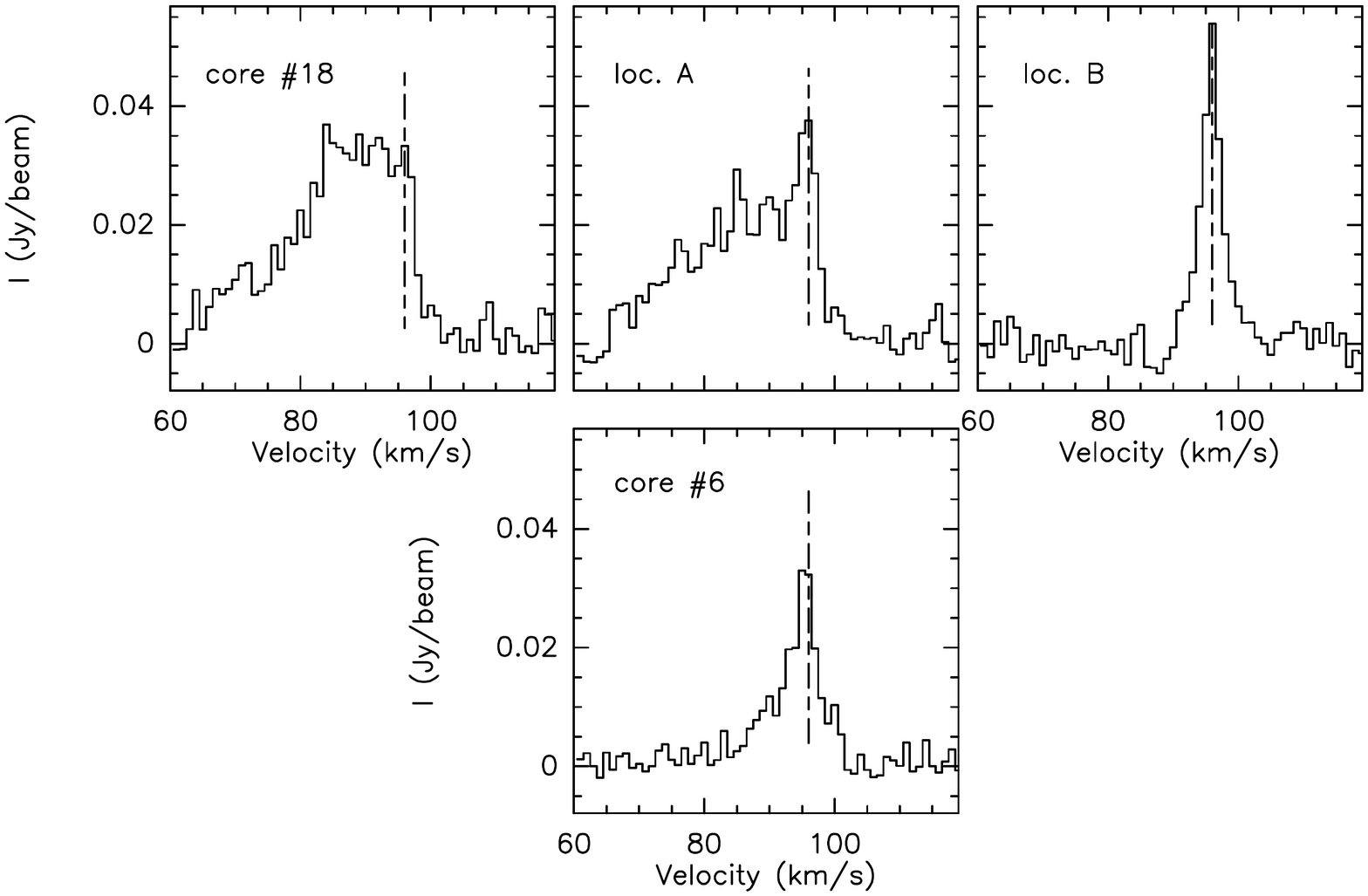}
%\vspace{-1.8cm}
\caption{Spectra of the SiO(5-4) line toward the continuum peak of cores \#6 and \#18 and toward locations A and B (see Fig.~\ref{co-s6}b and Sect.~\ref{s:SF}). The vertical dashed lines indicate the velocity of the SiO line at location B ($96~\rm km\,s^{-1}$).}
\label{siolines}
\end{figure*}

\end{appendix}

\end{document}